| | |
|---|---|
| Title | CO$_2$ splitting by DBD: understanding the influence of electrical parameters and regimes |
| Authors | A. Ozkan[1,2], T. Dufour[1], T. Silva[3], N. Britun[3], R. Snyders[3], A. Bogaerts[2] and F. Reniers[1] |
| Affiliations | [1] Université Libre de Bruxelles, Chimie analytique et chimie des interfaces (CHANI), Campus de la Plaine, Bâtiment A, CP255, boulevard du Triomphe, 1050 Bruxelles, Belgium<br>[2] Research group PLASMANT, Department of Chemistry, Universiteit Antwerpen, 2610 Antwerpen-Wilrijk, Belgium<br>[3] Université de Mons, Chimie des Interactions Plasma-Surface (ChIPS), CIRMAP, 23 Place du Parc, 7000 Mons, Belgium |
| Ref. | ISPC-22, Antwerp, Belgium, 5th-10th July 2015, P-II-8-22<br>http://www.ispc-conference.org/ispcproc/ispc22/P-II-8-22.pdf |
| DOI | |
| Abstract | Plasma processing is an innovative approach for the decomposition of CO$_2$ into O radicals and CO as a valuable carbon source. In this experimental work, a tubular dielectric barrier discharge operating at atmospheric pressure has been used to split CO$_2$ and to study its conversion considering the influence of frequency and power, as well as the influence of various electrical regimes (AC, AC pulsed regimes). The CO$_2$ conversion has been measured by mass spectrometry and gas chromatography, while the gas and wall temperatures have been determined and correlated to evaluate their influence of the CO$_2$ splitting. |

# Introduction

Carbon dioxide is among the most important greenhouse gases produced by industries and taking part to the global warming [1]. As a consequence, the conversion of this highly stable gas has gained in interest over the last decades, particularly its conversion into more value-added products such as carbon monoxide and oxygen [2]. In that framework, a plasma is a promising method to convert carbon dioxide [3]. Dielectric barrier discharges can overcome the CO$_2$ inertness via the production of electrons which activate the molecules, so that the splitting is stimulated by the vibrational excitation, while the ionization is stimulated by electronic excitation [4-6]. In this experimental work, we have investigated the influences of electrical parameters (frequency and power) and electrical regimes (AC and AC pulsed) on the CO$_2$ conversion, the discharge current, the filamentation and the gas temperature in order to understand the splitting mechanisms of CO$_2$.

# 2. Experimental setup

A cylindrical DBD reactor dedicated to the treatment of high flow rates of CO$_2$ gases has been designed as shown in Fig. 1a. It consists of a 2mm thick tube made of alumina with an inner electrode biased to the AC high voltage and an outer electrode which is grounded. The inner electrode is a copper rod with a diameter of 22 mm and a length of 180 mm while the outer electrode is a stainless steel mesh that can easily be rolled around the tubular dielectric barrier. The power applied to the high-voltage electrodes is provided by an AFS generator G10S-V with a maximum power of 1000 W. This generator can be coupled to a transformer with the ability to tune the frequency in the 1-30 kHz range. The gap, i.e., the distance separating the inner electrode from the dielectric barrier, is 2 mm, while the length of the discharge is 100 mm (so as to ensure a long residence time).

As sketched in Fig. 1b, the products resulting from the plasma phase reactions are analysed downstream of the reactor either with an online gas chromatograph (Agilent 7890) equipped with an Agilent column (40.0 µm 19095P-QO4) or by mass spectrometry (Hiden Analytical QGA). Optical emission spectroscopy (Andor Shamrock 500i) has also been carried out to determine the rotational temperature of the CO band and therefore the gas temperature [7]. Besides, the temperature of the reactor inner walls has also been measured using infrared imaging. Finally, electrical characterizations have been performed using an oscilloscope Tektronix DPO 3032 to measure the power absorbed by the plasma and the discharge currents.

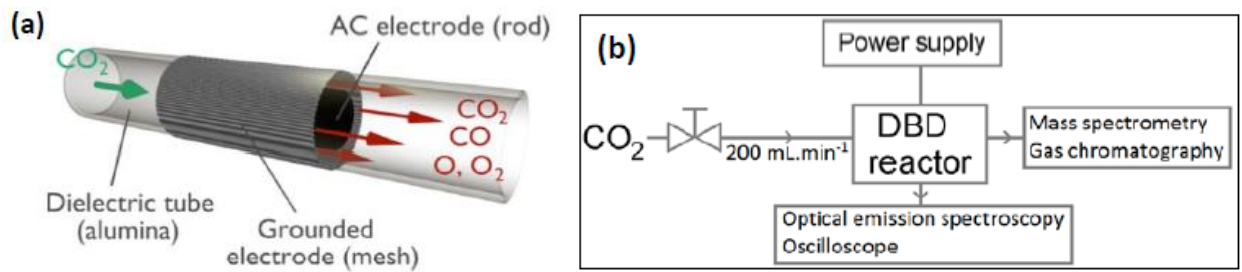

*Fig. 1. Schematic diagrams of (a) the DBD reactor, and (b) exp. setup.*

## 3. Results & Discussion

### 3.1. Frequency effect

The AC voltage has been applied to the inner electrode while the outer electrode has been grounded. The influence of the frequency on the $CO_2$ conversion has been investigated using two generators coupled with their respective transformers: first in a range comprised between 2.1 and 3.3 kHz and second in the 16.2-28.6 kHz range. As seen in Fig. 2, a quasi-linear decrease in the $CO_2$ conversion is obtained as a function of the frequency, in both cases. A conversion as high as 25% is reached at a frequency of 2.0 kHz, while a conversion of 17% is obtained at 28.6 kHz.

To understand how the frequency affects the $CO_2$ conversion, oscillograms of the discharge current have been realized and are reported in Fig. 3, considering three frequencies: 3.3 kHz, 16.2 kHz and 28.6 kHz. At 3.3 kHz, the current peaks show a maximum magnitude turning around 0.2 mA for positive voltages, and -1.5 mA on negative voltages. By comparison, the current peaks at 16.2 kHz and 28.6 kHz show maximum magnitudes of only |0.2 mA| whatever the sign of the voltage. The main difference between these two last frequencies is the number of current peaks, which is around 440 streamers/period at 16.2 kHz, and 360 streamers/period at 28.6 kHz. These values can also be expressed on the same residence time, for instance 120 µs: 1760 streamers at 16.2 kHz and 2160 streamers at 28.6 kHz. As a result, even if the filamentation of the discharge is stronger at high frequency, it does not seem to induce a higher $CO_2$ conversion. However, this rate may be enhanced by the discharge temperature, as discussed afterwards.

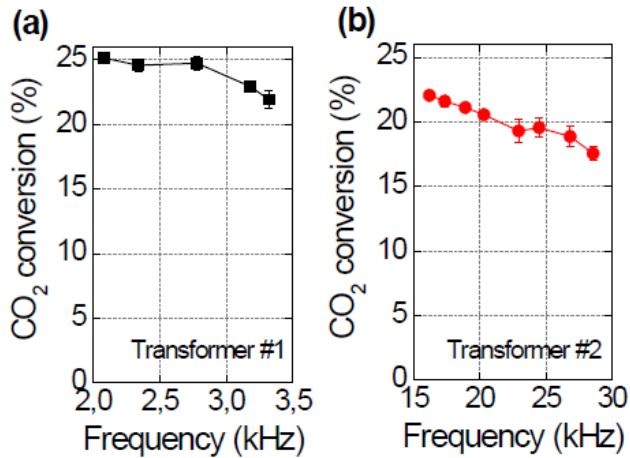

Fig. 2. $CO_2$ conversion as a function of the frequency, for P = 60 W and $CO_2$ flow rate = 200 $mL_n$/min.

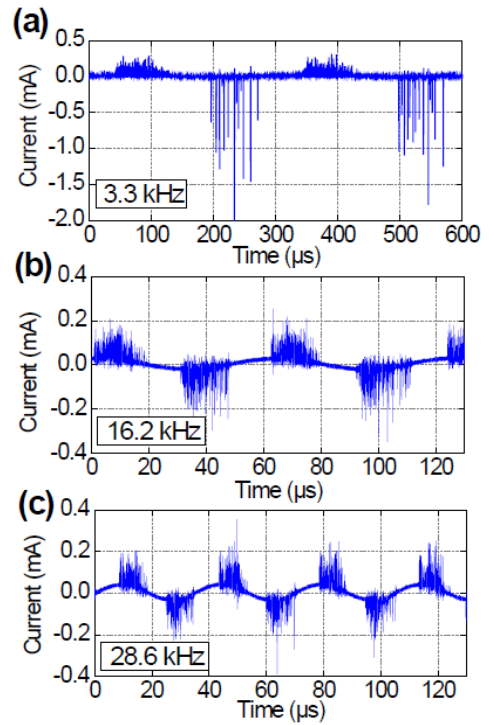

Fig. 3. Current oscillograms for a $CO_2$ discharge ignited at 60 W for 3.3, 16.2 and 28.6 kHz.

We have also studied the effect of the frequency on the gas temperature. For this, the emission of the CO Angstrom band (bandhead at 483.3 nm) has been detected by optical emission spectroscopy and its ro-vibrational structure has been fitted to determine the rotational temperature (see Fig. 4). In our conditions, we show that Trot is very close to the temperature of the heavy molecules (CO) and therefore to the gas temperature of the discharge.

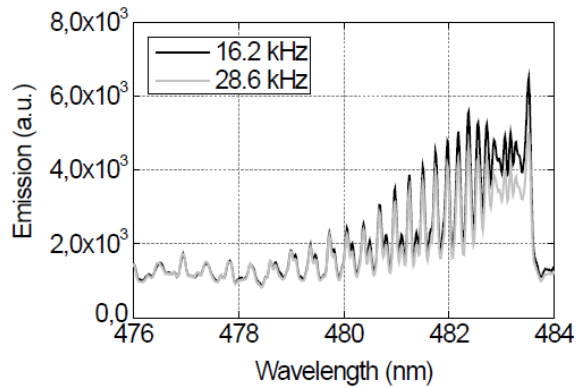

Fig. 4. CO band from the Angstrom System {$B^1\Sigma^+ \rightarrow A^1\Pi$ ; v'=0 → v''=1} for the $T_{gas}$ determination in $CO_2$ discharge at 2 different frequencies.

Finally, a correlation of these gas temperatures with the temperature of the DBD inner walls (Twalls) has been achieved. The value of Twalls has been determined using an infrared camera which allows the imaging of Twalls through two dimensional profiles, as shown in Fig. 5 for 0, 2, 4 and 6 minutes of $CO_2$ splitting treatment. A decay of the frequency induces a lower value for Twalls. For instance at 6 minutes, a wall temperature of almost 90 °C is obtained at 3.3 kHz against 140 °C at 28.6 kHz. Twalls can be correlated with Tgas to evaluate the plasma heating losses (Joule effect) and their effect on the $CO_2$ conversion. Furthermore, operating at low frequency seems to improve the spatial uniformity of the discharge throughout the entire interelectrode region because of the heat losses spread over this entire region. On the contrary, increasing the frequency (28.6 kHz) induces a partial confinement of the discharge, i.e., to approximately half of the interelectrode region. Such a confinement of the discharge may explain why at higher frequency (28.6 kHz), the $CO_2$ conversion is 17% while as mentioned in Fig. 2, it is almost 22% at 16.2 kHz (and even higher at 3.3 kHz).

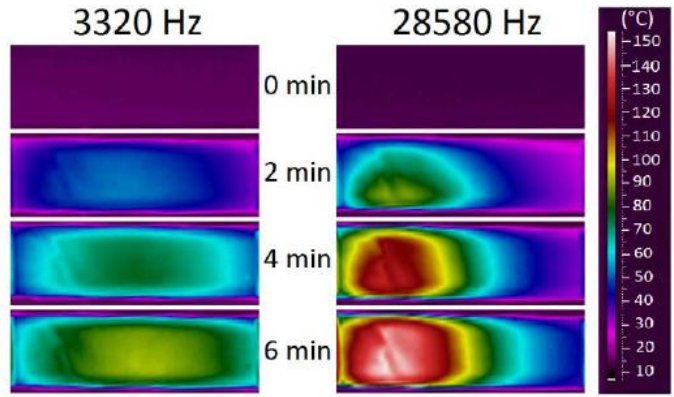

Fig. 5. Infrared images of the discharge zone as a function of time for two different frequencies.

## 3.2. Influence of the electrical regime

The influence of the electrical regime has been studied by comparing pure AC and pulsed AC voltages on the $CO_2$ conversion. In both cases, the relative power applied by the generator has been set to 50 W, which means that the product of the pulsed power by the duty cycle remains constant during the experiments. In the pulsed AC regime, pulses have been applied always keeping a same width of 1 ms and increasing the duty cycle from 40% to 100% so as to decrease the peak power from 110 W to 50 W, as reported in Fig. 6. The $CO_2$ conversion is much higher for low duty cycles: for a pure AC regime (duty cycle = 100%), it is 16% while it reaches a value as high as 26% for a duty cycle of 40%. In other words, the $CO_2$ conversion obtained in pulsed AC regime at high frequency (28.6 kHz) is almost the same as the one obtained in pure AC regime at low frequency (3.3 kHz).

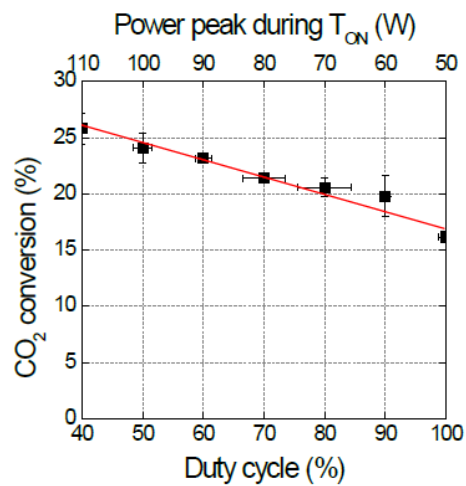

Fig. 6. $CO_2$ conversion as a function of duty cycle for f = 28.6 kHz and $CO_2$ flow rate = 200 mL$_n$/min.

# Acknowledgement

This work is supported by PSI-IAP 7 (Plasma Surface Interactions) from the Belgian Federal Government BELSPO agency.